\documentstyle[osa,manuscript,psfig]{revtex} 

\begin{document}
\titlepage

\title{Electron as Spatiotemporal Complexity due to Self-Organized Criticality}

\author{Meng Ta-chung\protect\\
{\em Institute of Particle Physics, CCNU, 430079 Wuhan, China}\\
{\em Institut f\"ur Theoretische Physik, FU-Berlin,
14195 Berlin, Germany}\protect}

\maketitle

\begin{abstract}
  The electron, which has been pictured as an elementary particle ever
  since J.J. Thomson's $e/m$-measurement in 1897, and the relativistic
  motion of which is described by the Dirac equation, is discussed in
  the light of the recent progress made in Science of Complex Systems.
  Theoretical arguments and experimental evidences are presented which
  show that such an electron exhibits characteristic properties of
  spatiotemporal complexities due to Self-Organized Criticality (SOC).
  This implies in particular that, conceptually and logically, it is
  neither possible nor meaningful to identify such an object with an
  ordinary particle, which by definition is something that has a fixed
  mass (size), a fixed lifetime, and a fixed structure.
\end{abstract}


The electron has been pictured as a particle, and as one of the
elementary building blocks of nature, ever since J.J. Thomson
published the result of his $e/m$-measurement in 1897 \cite{thomson}.
The Dirac equation \cite{dirac2a,dirac2b,dirac}, originally designed
as an one-particle equation, which describes the relativistic motion
of such an electron ($e$) in the framework of Quantum Theory, is
undoubtedly one of the greatest achievements in physics --- although
the original goal failed. This is because, it is Dirac's equation
which predicted the existence of positron and thus led to the
discovery of one of the general fundamental symmetries (the
charge-conjugation symmetry) in nature \cite{dirac2b,dirac}; and
because, it is also this equation which describes in general the
relativistic motion of all the known (electrically charged) spin-$1/2$
objects, in particular, that of the heavier leptons ($\mu$, $\tau$),
as well as that of the quarks ($u$, $d$, $s$, etc) which are the
constituents of hadronic matter \cite{Bjorkendrell,gottfried}. The
purpose of this letter is to point out that the above-mentioned
electron exhibits characteristic properties of spatiotemporal
complexities due to self-organized criticality (SOC)
\cite{BTWoriginal,BTWcontinue,GluonSOC,Jet}.  Hence, among other
things, it is neither possible nor meaningful to identify such an
object with an ordinary particle which, by definition, should have a
fixed size (mass), a fixed lifetime, and a fixed structure. The
theoretical arguments and experimental evidences which lead to this
conclusion can be summarized as follows: Two of the fundamental
symmetries in nature, namely the invariance under charge conjugation
transformation and that under (Abelian) local gauge transformation,
which is also known as the gauge principle \cite{gottfried}, dictate
the following: The validity of Dirac's equation for the spin-$1/2$
charged fermions $e$, $\mu$, ..., $s$, $c$, etc.  implies not only the
existence of the corresponding antifermions $\bar{e}$, $\bar{\mu}$,
..., $\bar{s}$, $\bar{c}$, etc., but also that the pairs made out of
the above-mentioned fermions and antifermions interact with the
electromagnetic field in one and the same manner.  This means in
particular that , electromagnetic interaction can produce transitory
virtual pairs $e^+-e^-$, $\mu^+-\mu^-$, ..., $s-\bar{s}$, $c-\bar{c}$,
etc. in the same ``vacuum''.  Taken together with the fact that all
Dirac fermion fields fluctuate
\cite{dirac2a,dirac2b,dirac,dirac3,Bjorkendrell,gottfried}, the
picture we obtain for ``the electron'', by examining it in a spatial
region within its Compton wavelength, is the following: It is always
closely associated with --- in fact inseparable from --- ``the
vacuum'' which consists of an indefinite number of all kinds of
transitory virtual fermion-antifermion pairs ($e^+-e^-$,
$\mu^+-\mu^-$, ..., $s-\bar{s}$, $c-\bar{c}$, etc.  with various
indefinite lifetimes) where the total number, as well as the relative
abundance, of such pairs are in general different in different regions
of space-time. In other words, ``the electron'' has to be considered
as an open dynamical complex system of an indefinite number of degrees
of freedom.  Precisely speaking, ``the electron'' is a typical
spatiotemporal complexity due to SOC
\cite{BTWoriginal,BTWcontinue,GluonSOC,Jet}. The proposed picture, in
which ``the fluctuating vacuum polarization'' deduced from Dirac's
equation \cite{dirac2a,dirac2b,dirac,dirac3,Bjorkendrell,gottfried} is
viewed as an integral part of nature, can be readily tested by
comparing it with the available electron-positron collision data
\cite{data1,data2,data3}. The results of such tests strongly support
the picture.

It has been pointed out some time ago by Bak, Tang and Wiesenfeld
\cite{BTWoriginal,BTWcontinue} that striking simple regularities exist
in many seemingly disparated open dynamical complex systems far from
equilibrium. Based on the fact that the same regularities appear in
many seemingly not comparable systems in the macroscopic world ---
from the formation of the landscape to the process of evolution to the
action of nervous systems to the behaviour of the economy --- it is
natural to ask \cite{GluonSOC,Jet} whether such regularities are so
general and so universal that they are true also in the microscopic
world. The questions we discuss in this paper are the following.  Down
to what level can we find such open dynamical complex systems? What is
an ``electron'', the motion of which is described by the Dirac
equation? Can such an ``electron'' indeed be viewed in the way we
usually do in the conventional picture (see below for more details),
namely as an ordinary particle which should and can have a fixed mass,
a fixed lifetime, and a fixed (in particular point-like) structure?

We recall that the picture of ``the electron'' (it will hereafter be
referred to as ``the conventional picture'') in the renormalized
Quantum Electrodynamics (QED) which is based on the validity of Dirac
equation for ``the electron'' and the validity of Maxwell equations
for the electromagnetic field) can be described as follows
\cite{Bjorkendrell,gottfried}: Similar to the corresponding picture in
Classical Electrodynamics, ``the electron'' is, also in this case, a
stable particle, because no ``excited electron'' and no
``electron-decay'' has been observed experimentally. Its mass ($m_e$)
and its charge ($e$) have fixed values; and these values can be, and
have been, determined (by measuring $e/m_e$ and $m_e/m_H$ where $m_H$
is the mass for the hydrogen atom, or by measuring ``the elementary
charge'' a la Millikan) by performing experiments at low velocities,
which means that they are measured at spatial distances large compared
to the electron Compton wavelength $1/m_e$. But, ``the electron''
which is loosely called ``a point charge in vacuum'' must, on the
other hand, be envisaged as being surrounded by an induced charge
distribution that steems from transient $e^+-e^-$ pairs produced by
the electromagnetic field of ``this point charge''.  Precisely
speaking, according to perturbative QED calculations, distribution of
the induced charges extends out to distances of order $1/m_e$, and is
a small and diffuse effect except at distances vastly shorter than
those currently attainable by experiment. Note that this conclusion is
reached under the assumption that the Fourier transform of a static
charge distribution (known as its ``electromagnetic form factor'') can
be used to describe the electromagnetic structure of a spatially
extended object, not only in the non-relativistic limit of slowly
moving particles, but also in describing those in relativistic
scattering processes. In fact, ``the size'' of the charged lepton ($e,
\mu, \tau$) has been determined by examining the deviations of their
electromagnetic form factors from unity (by definition
``structureless'') in high-energy collision processes such as $e^+ +
e^-\rightarrow \mu^+ + \mu^-$, $e^+ + e^-\rightarrow e^+ + e^-$.
These, as well as other related facts show that, when we use the
conventional picture for ``the electron'', it is of considerable
importance to keep the following in mind: The claim that the leptons
are point-like {\em stems from theory}. The observation of an object's
structure {\em involves not only a probe, but also a detailed
  knowledge of that probe's interaction with the object}. In the
conventional picture, the probe is electromagnetic field, and the
probe's interaction with the object is described by QED and other
related theories/models in particular the above-mentioned assumption
about electromagnetic form factors. Here, it is of particular
importance to recall that, due to electromagnetic interaction, charges
radiate, and such radiation also interacts with the emitting charges.
By taking these effects into account, perturbative QED calculations
yield a diverging expression for ``the self-energy'' which contributes
to the physical mass, and a diverging expression for ``the induced
charge'' due to ``vacuum polarization''.  Finite results are obtained
by performing the renormalization procedure in the following way:
``Renormalized quantities'', ``the renormalized mass'' and ``the
renormalized charge'' (also known as ``the dressed charge''), are
introduced into QED. They are defined as the differences between the
above-mentioned diverging expressions and the quantities which
originally appear in the Dirac equation for ``the electron'' in
electromagnetic field . The original mass and charge are called ``the
bare mass'' and ``the bare charge'' respectively, and both of them are
assumed to be infinite. The ``renormalized quantities'' are finite;
their values are obtained by identifying them with the corresponding
measured quantities in low-velocity experiments where the measurements
are made at distances larger than the electron Compton wavelength
$1/m_e$ .  This means, not only ``the point-like structure'', but also
``the mass'', ``the charge'', and ``the lifetime'' of ``the electron''
in the conventional picture are {\em theoretical inputs} which are
{\em additional rules} set up in a self-consistent way to avoid the
infinities originate from ``the fluctuating sea of negative-energy
electrons'' in the Dirac theory.

In short, the conventional picture for ``the electron'' in the
framework of QED is obtained through a straightforward modification:
The concepts and methods which have been used to describe a charged
particle in classical physics (mechanics and electrodynamics) are
replaced by the corresponding ones in quantum physics (quantum
mechanics and quantum field theory). But, such a modification leads to
{\em conceptual, logical}, as well as {\em computational}
difficulties, and almost all of these difficulties are intimately
related to the properties of the Dirac equation. The {\em
  computational} difficulties have been circumvented by introducing
``the renormalized mass'', ``renormalized charge'' etc. for ``the
electron''. While good agreement between the measured results and the
above-mentioned theoretical inputs has been achieved, {\em the
  difficulties} caused by the fluctuations of ``the negative-energy
electrons'' in the Dirac theory {\em have not been removed}.

The importance of the above-mentioned {\em conceptual} and {\em
  logical} difficulties in the Dirac theory {\em cannot} be
exaggerated. Due to the fact that QED is based on Dirac's equation on
the one hand, and the fact that ``the electron'' {\em is assumed} to
be a structureless point-like particle is in sharp contrast to the
nature and the spirit of the Dirac equation on the other hand, it
isnot difficult to understand why many concerned physicists {\em
  cannot} accept the idea that the present form of QED (and/or that of
the Electroweak Theory) should already be {\em the final version} of a
quantum theory for the relativistic motion of an electron in
electromagnetic field. In this connection, it is perhaps of some
interest to recall what Dirac wrote --- as the last sentence --- in
his book published in 1958 \cite{dirac}: ``The difficulties, being of a
profound character, can be removed only by some drastic change in the
foundations of the theory, probably a change as drastic as the passage
from Bohr's orbit theory to the present quantum mechanics.''

Due to the tremendous success --- mainly caused by the existence of
``negative energy solutions'' --- the Dirac equation {\em has not
  been} and {\em should not be} abandoned, although it is known almost
immediately after its discovery
\cite{dirac2a,dirac2b,dirac,dirac3,Bjorkendrell,gottfried}, that it {\em
  cannot} be an one-particle equation. But in QED, ``the electron'' is
by definition a particle with a fixed mass and a fixed structure, the
motion of which is described by Dirac's equation. Hence, as a first
step towards a possible removal of the conceptual and logical
difficulties, where a ``drastic change'' cannot be avoided, it is
perhaps of some interest to find out the following: Are there {\em
  compelling reasons} which {\em force} us to accept the theoretical
input, that ``the electron'' {\em should} be viewed as an ordinary particle
with a fixed mass, a fixed lifetime, and a fixed structure? If not,
why can we not abandon the additional theoretical inputs, and accept
the logical consequences of the Dirac theory and those of the gauge
principle? To be more precise, why can we not accept that the virtual
fermion-antifermion pairs in ``the vacuum'' indeed play a significant
role --- especially when ``the electron'' is localized within its
Compton wavelength? Why can we not simply accept that ``the electron''
is inseparable froman indefinite number of fermion-antifermion pairs
which are fluctuating all the time and everywhere? In other words, why
do we not simply accept that ``the electron'' should be considered as
an open dynamical complex system?

Once we agree to accept this, the next question we have to face is to
find out whether it is possible to probe such open dynamical complex
systems experimentally --- especially to probe ``the interior'' of
such a system withinthe electron Compton wave length. Besides the
well-known fact \cite{gottfried} that the effects of ``negative energy
solutions'' of the Dirac equation are particularly significant in this
region, there are also other reasons why this region is of particular
interest: (I) The validity of the uncertainty principle and the
special relativity theory dictates that the distance between the
members of a virtual pair due to vacuum polarization cannot exceed the
corresponding Compton wavelength. This implies in particular that the
distances between the members of a $s-\bar{s}$ or a $c-\bar{c}$ pair
are much shorter than that between the members of an $e^+ -e^-$ pair,
and thus it is expected that the effects on charge distribution caused
by the former can be detected only when the distance between the probe
and the object under investigation is much smaller than the electron
Compton wavelength.  (II) It is within this region, where the
theoretical inputs in the conventional picture have been used to
circumvent the diverging results obtained by performing perturbative
QED calculations.

Can we find {\em experimental} means to probe ``the {\em interior}''
of ``the electron'' ---without making use of the above-mentioned
theoretical inputs? To be more precise, can we answer the following
questions? (A) Are there experiments in which we can see that, not
only virtual $e^+-e^-$ pairs, but also {\em virtual pairs of heavy
  objects} (e.g. $s-\bar{s}$ and $c-\bar{c}$) {\em exist} in ``the
vacuum'' which is intimately associated with, in fact inseparable
from, ``the electron''? (B) Are there experiments in which we can see
that such created virtual $s-\bar{s}$ and $c-\bar{c}$ pairs of ``the
electron'' can ``disappear'' in the sense thatsuch heavy pairs turn
into radiations and thus ``return to the vacuum''? In other words, are
there experimental evidences that such heavy quark-antiquark pairs can
turn into something which behave very much the same as electromagnetic
radiation? (C) Are there experiments which show how the emergence and
the extinction of such virtual fermion-antifermion pairs take place?
Is it possible to extract information about the emergence-extinction
processes from the existing experimental data?

As can be seen in more detail below, a simple and convenient way to
examine ``the interior'' of ``the electron'' and/or that of ``the
positron'' is to letthese two objects get as close as possible to each
other in space-time such that we can also examine the various effects
caused by energetic quark-antiquark pairs. Here, we make use of the
following (empirical, actually rather trivial) fact.  The resulting
system formed by two open dynamical complex systems with the same kind
of ingredients is again an open, dynamical, and complex one. Under
normal conditions (e.g. not at certain ``resonance energies'' where
certain special reactions may take placed), it is expected that their
characteristic properties also remain unchanged. For example,the
resultant system formed by two systems far from equilibrium remains to
be far from equilibrium.  Having this in mind, we now look at the
available data \cite{gottfried,data1,data2,data3} obtained from high-energy
electron-positron collision experiments. To be more specific:

(A) We examine the characteristic features of the well-known
\cite{gottfried,data1} ratio $R(w) = \sigma_T(e^++e^-\rightarrow
hadrons; w) / \sigma_{\mu\mu}(e^++e^-\rightarrow \mu^++\mu^-; w)$,
where $\sigma_T$, $\sigma_{\mu\mu}$, and $w$ stand for the total
cross-section, the integrated cross-section for $e^++e^-\rightarrow
\mu^++\mu^-$, and the total c. m.  s.  energy of the $e^++e^-$ system
respectively; and we examine in those two kinds of reactions the
characteristic features of the angular distributions of the produced
hadrons. The observation \cite{gottfried,data1,data2} that $R(w)$ appears
approximately as ascending steps for increasing $w$, where every jump
corresponds to a threshold at which the creation of a quark-antiquark
pair of a given flavour becomes possible; as well as the observation
\cite{data2} that the produced hadrons appear mainly in two jets where
the forward- and the backward-directions are preferred, show that the
question raised in (A) should be answered in the affirmative.  In
other words, these experimental facts show that there are not only
virtual $e^++e^-$ pairs but also virtual heavy objects such as
quark-antiquark pairs, distributed in the region within the electron
Compton wavelength of the colliding electron/positron. The reasons
which lead us to this conclusion are the following: Hadrons are
colour-singlets, and the overwhelming majority of the produced hadrons
are flavour-neutral which can be formed only when suitable energetic
quarks and antiquarks associated with the colliding electron/positron
can meet and interact with each other. At sufficiently high incident
energies, the members of such fermion-antifermion (i.e. either
lepton-antilepton or quark-antiquark) pairs are in general energetic,
and their momenta are mainly in the direction of the colliding
electron/positron. Every quark (or antiquark) has the chance to meet
another suitable partner to become a hadron. Virtual quark-antiquark
pairs such as $s-\bar{s}$ and $c-\bar{c}$ may also become physical
(real) by acquiring energy from other virtualpairs (of any flavour).
Forward- and backward-directions are preferred, because of momentum
conservation. As $w$ increases, more and more varieties of heavier
quarks/antiquarks can be created, and thus the chance for the
production of energetic and physical heavy quark-antiquark pairs
becomes larger. We note that pairs of heavy quark-antiquark may also
appear as meson-resonances with various quantum numbers at
resonance-energies.

(B) We examine the yield of $\mu^+-\mu^-$ pairs in the final state of
the reaction $e^++e^-\rightarrow \mu^++\mu^-$, and look for
interference effects in the neighbourhood of the above-mentioned
thresholds. The fact that spectacular interference effects have been
observed \cite{data1} in the neighbourhood of the $f$- and in that of
the $J/\psi$-resonances show that the questions raised in (B) should
be answered in the affirmative.  This is because, as mentioned in (A),
the existence of virtual quark-antiquark pairs in ``the electron'' and
``the positron'' implies, that at sufficiently high incident energies,
virtual as well as real quark-antiquark pairs --- including
meson-resonances --- can be formed, and some of them (any sort of
them, in particular the resonances) may have the same quantum numbers
as those of the photon.  In terms of the usual quantum mechanical
description, the amplitude of a resonance is complex --- in contrast
to that of a photon which is real. Hence, measurable interference
effects are expected when the above-mentioned process takes place
nearthe resonance-energies. We note that the reason for examining such
effects in the $e^++e^-\rightarrow \mu^++\mu^-$ rather than in the
$e^++e^-\rightarrow e^++e^-$ channel is that although in both cases,
the intermediate state can have quantum numbers same as the photon,
but only in the former case, no identical particle and/or
``cross-channel contribution'' effects need to be taken into account.
We also note, the observation \cite{data1} of such interference
effects explicitly shows that not only lepton-antilepton pairs but
also quark-antiquark pairs disappear in the same way as they emerge in
``the vacuum'' which is an integral part of ``the electron'' and/or
``the positron'' described by the Dirac equation(s)!

\begin{figure*}
  \psfig{figure=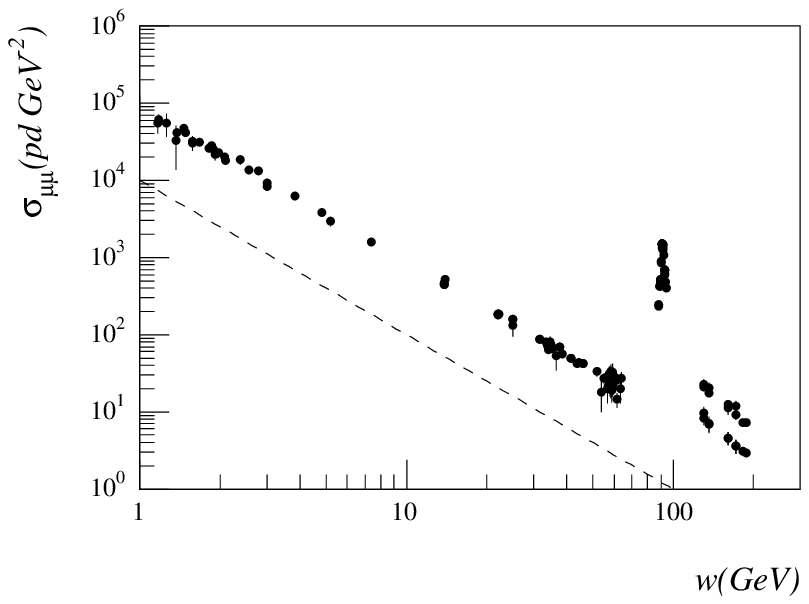}
  \caption{The integrated cross-section $\sigma_{\mu\mu}$ for the 
    reaction $e^++e^-\rightarrow \mu^++\mu^-$ is plotted as function
    of the total c.m.s energy $w$ of the $e^+-e^-$ system. The
    existing data, taken from Refs.[\ref{data1}-\ref{data2}], lie
    approximately on one straight line in this log-log plot. The
    dashed line is drawn to show that the slope of the is line is -2,
    and thus it is exactly the same as Raup's ``kill-curve'' for
    biological evolution [\ref{raup},\ref{BTWcontinue}].}
  \label{fig}
\end{figure*}

(C) We examine the energy ($w$)-dependence of the integrated
cross-section ($\sigma_{\mu\mu}$) for the reaction $e^++e^-\rightarrow
\mu^++\mu^-$ at sufficiently high total c. m. s.  energies ($w$'s).
The fact that almost all (as expected, the only exceptions being those
in the vicinity of the weak gauge boson $Z$.) the available data
points \cite{data3} lie on one straight line with a slope -2 in the
log $\sigma_{\mu\mu}$ vs. log $w$ plot (cf.  Figure \ref{fig}) show
that also the questions in (C) should be answered in the affirmative.
It is because this plot shows how the lifetimes of such virtual pairs
are distributed. Note in particular that, in general, such virtual
pairs are already present either as ``part of the electron'' or ``part
of the positron'' before the two open dynamical complex systems come
close (of the order of electron Compton wavelength, say) to each
other, and we examine those extinction processes which take place in
the resultant open system, when members of high-energy virtual pairs
find their suitable high-energy suitable partners and thus ``return to
the vacuum''.  To be more specific, the plot clearly shows the
following: First, there are strong experimental indications that the
ingredients of ``the electron'' are changing all the time and thus it
is neither meaningful nor possible to say that ``the electron'' is a
particle with a fixed mass, a fixed lifetime, and a fixed structure.
Second, the evolution of the various virtual pairs in ``the electron''
obeys the same law as the kill-curve discovered by Raup
\cite{raup,BTWcontinue} for biological evolution --- a particularly
striking fingerprint of self-organized criticality SOC
\cite{BTWoriginal,BTWcontinue,GluonSOC,Jet}. 

In order to see explicitly how the above-mentioned conclusions have
been reached, it is useful to keep the following in mind.

(I). Fermions and antifermions can in general recombine.  Although the
detailed interactions between fermions and antifermions as well as
those between the pairs are in general different, such details do not
play a significant role as far as thelifetime distribution is
concerned, provided that they are constituents of an open dynamical
complex system due to SOC \cite{BTWoriginal,BTWcontinue,GluonSOC,Jet}.
(II). The lifetime ($\tau$) of virtual constituents of a system ---
such as ``the electron'' or the resultant system --- can be estimated
by making use of the uncertainty principle and the basic ideas of
Feynman's parton model \cite{data2}. The result, which is now
well-known \cite{data2}, is that, in the c. m. s. frame of ``the
electron'' and ``the positron'', the lifetime of a virtual object in
either of them is directly proportional to the magnitude of the total
longitudinal momentum of either one of the colliding objects, and thus
it is directly proportional to $w$, the total c. m. s energy of the
above-mentioned colliding system. Hence, the $\tau$-dependence of the
probability for a virtual pair of fermion-antifermion to disappear,
that is ``back to vacuum'', by turning into a $\mu^+-\mu^-$ (either
through recombination or through acquiring energy from other pairs) is
directly proportional to the $w$-dependence of this probability.We note
that the latter is nothing else but the integrated cross-section ($w$)
for $e^++e^-\rightarrow \mu^++\mu^-$.

In conclusion, not only theoretical arguments, but also the existing
high-energy electron-positron collision data \cite{data1,data2,data3},
strongly suggest that ``the electron'' described by the Dirac equation
\cite{dirac2a,dirac2b,dirac} is a spatiotemporal complexity due to
self-organized complexity \cite{BTWoriginal,BTWcontinue,GluonSOC,Jet}
. It is indeed a remarkable fact that the emergence and extinction of
various types of virtual fermion-antifermion pairs in ``the electron''
and the emergence and extinction of various genera on earth obey one
and the same law \cite{raup,BTWcontinue}!

The author thanks Liu Qin in Wuhan for the relevant question she asked
in the Relativistic Quantum Mechanics class. He thanks J. Meng, L.
Meng and K. Tabelow in Berlin for their valuable comments, and for
their help in performing the data analysis as well as in doing the
computer work.

\begin{thebibliography}{aaa}

\bibitem{thomson}\label{thomson}
  J. Y. Thomson, Phil. Mag. {\bf 44}, 293 (1897).

\bibitem{dirac2a}\label{dirac2a}
  P. A. M. Dirac, Proc. Roy. Soc (London) A {\bf 117}, 618 (1928).

\bibitem{dirac2b}\label{dirac2b}
  P. A. M. Dirac, Proc. Roy. Soc (London) A {\bf 118}, 315 (1928).

\bibitem{dirac}\label{dirac}
  P. A. M. Dirac, {\em The Principles of Quantum Mechanics}, (Oxford
  University Press, London, 1958), 4th ed.

\bibitem{Bjorkendrell}\label{bjorkendrell}
  See e.g. J.D. Bjorken and S.D. Drell, {\em Relativistic Quantum
    Mechanics}, (McGraw-Hill, 1964), p. 38.

\bibitem{gottfried}\label{gottfried}
  See e.g. K. Gottfried and V. F. Weisskopf, {\em Concepts of Particle
    Physics}, (Oxford University Press, 1986).

\bibitem{BTWoriginal}\label{BTWoriginal}
  P. Bak, C. Tang, and K. Wiesenfeld, Phys. Rev. Lett. {\bf 59}, 381
  (1987); Phys. Rev. A {\bf 38}, 364 (1988).

\bibitem{BTWcontinue}\label{BTWcontinue}
  P. Bak, {\em How Nature Works}, (Springer, New York, 1996).

\bibitem{GluonSOC}\label{GluonSOC}
  T. Meng, R. Rittel, and Y. Zhang, Phys. Rev. Lett. {\bf 82}, 2044
  (1999); C. Boros, T. Meng, R. Rittel, K. Tabelow, and Y. Zhang,
  Phys. Rev. D {\bf 61}, 094010 (2000).

\bibitem{Jet}\label{Jet}
  J. Fu, T. Meng, R. Rittel, and K. Tabelow, Phys. Rev. Lett. {\bf
    86}, 1961 (2001).

\bibitem{dirac3}\label{dirac3}
  O. Klein, Z. Phys. {\bf 53}, 157 (1929); E. Schr\"odinger, Sitzber.
  Preuss. Akad. Wiss. Physik-Math., {\bf 24}, 418 (1930); P. A. M.
  Dirac, Proc. Roy. Soc. (London) A {\bf 126}, 360 (1930); J. R.
  Oppenheimer, Phys. Rev. {\em 35}, 939 (1930).

\bibitem{data1}\label{data1}
  See e.g. R. F. Schwitters and K. Strauch, Ann. Rev. Nucl. Sci. {\bf
    26}, 89 (1976); S. L. Wu, Phys. Reports {\bf 107}, nos.
  2-5 (1984) and the papers cited therein.

\bibitem{data2}\label{data2}
  See e.g. V. D. Baager and R. J. Phillips, {\em Collider Physics},
  (Addison Wesley, 1987) and the papers cited therein.

\bibitem{data3}\label{data3}
  W. Bartel, Z. Phys. C {\bf 30}, 371 (1986); W. Braunschweig, Z. Phys.
  {\em C40}, 163 (1988); B. Howell, Phys. Lett. B {\bf 291}, 206
  (1992); D. Buskulic, Phys. Lett. B {\bf 378}, 373 (1996); M.
  Acciarri, Phys. Lett. B {\bf 407}, 361 (1979); M. Acciari, Phys.
  Lett. B {\bf 479}, 101 (2000).

\bibitem{raup}\label{raup}
  D. M. Raup, Science {\bf 231}, 1528 (1986).





  
\end{thebibliography}
\end{document}